\documentclass[aps,prd,reprint,nofootinbib]{revtex4-2}

\usepackage{amsmath}
\usepackage{amssymb}	% Math symbols such as \mathbb
\usepackage{bbold}
\usepackage{bm} %allows msans to be used (for matrices' lettering)

\usepackage{xcolor}
\definecolor{myBlue}{RGB}{50,117,180}
\definecolor{myRed}{RGB}{200,40,40}
\definecolor{myGreen}{RGB}{34,139,34}
\definecolor{myLightBlue}{RGB}{193,223,255}
\definecolor{myOrange}{RGB}{255,185,0}

\usepackage{cancel} % cancel terms out of an equation with \cancel{} or \cancelto{}{}
\usepackage[normalem]{ulem} %allows \sout{} for strikethrough characters
\usepackage{changepage}
\usepackage{float}
\usepackage{mathrsfs}
\usepackage{url}
\usepackage{xfrac}
\usepackage{graphicx}
\usepackage{lipsum} %Lorem ipsum
\graphicspath{{/}}
\newcommand{\comment}[1]{\ignorespaces} %for inline commenting
\renewcommand{\v}[1]{\ensuremath{\mathbf{#1}}} % for vectors
 
\DeclareMathAlphabet{\mathsfit}{\encodingdefault}{\sfdefault}{m}{n}
\SetMathAlphabet{\mathsfit}{bold}{\encodingdefault}{\sfdefault}{bx}{n}
\newcommand{\ms}[1]{\bm{\mathsfit{#1}}} %block sans serif letters for matrices
\newcommand{\abs}[1]{\left| #1 \right|} % for absolute value
\makeatletter
\newcommand*\bigcdot{\mathpalette\bigcdot@{.5}}
\newcommand*\bigcdot@[2]{\mathbin{\vcenter{\hbox{\scalebox{#2}{$\m@th#1\bullet$}}}}}
\makeatother

\makeatletter
\newcommand{\raisemath}[1]{\mathpalette{\raisem@th{#1}}}
\newcommand{\raisem@th}[3]{\raisebox{#1}{$#2#3$}}
\makeatother

\usepackage{accents}

\usepackage{stackengine} %Allows \mathrel{\stackon[1pt]{$=$}{$\scriptstyle!$}}

\makeatletter %makes "@" just a letter
\renewcommand*\env@matrix[1][*\c@MaxMatrixCols c]{%
  \hskip -\arraycolsep
  \let\@ifnextchar\new@ifnextchar
  \array{#1}}
\makeatother

\usepackage{pgfplots}
\usepgfplotslibrary{patchplots}
\usepgfplotslibrary{external}
\usepackage{tikz}
\usetikzlibrary{external}
\usetikzlibrary{calc}
\usetikzlibrary{arrows,shapes,backgrounds,arrows.meta,shapes.arrows,automata,quotes,intersections}
\usetikzlibrary{decorations.text}
\usetikzlibrary{decorations.fractals}
\usetikzlibrary{decorations.pathmorphing}
\usetikzlibrary{decorations.markings}

\usetikzlibrary{decorations.pathreplacing}
\makeatletter
    \let\pgf@decorate@@brace@brace@code@old\pgf@decorate@@brace@brace@code
    \def\pgf@decorate@@brace@brace@code{
        \ifdim\pgfdecoratedremainingdistance<4\pgfdecorationsegmentamplitude
            \pgftransformxscale{\pgfdecoratedremainingdistance/4\pgfdecorationsegmentamplitude}
            \pgfdecoratedremainingdistance=4\pgfdecorationsegmentamplitude
        \fi
        \pgf@decorate@@brace@brace@code@old
    }
\makeatother

\newcommand*\mR{\mathbb{R}}

\newcommand*\mmP{\mathcal{P}}

\newcommand*\mD{\mathrm{D}}

\newcommand*\mL{\mathcal{L}}
\newcommand*\mO{\mathcal{O}}
\newcommand*\mS{\mathcal{S}}

\newcommand*\mmp{\mathfrak{p}}
\newcommand*\mso{{\mathfrak{so}(3,1)}}

\newcommand*\mOne{\mathbb{1}}

\newcommand*\md{\mathrm{d}}

\DeclareMathAlphabet{\mathpzc}{OT1}{pzc}{m}{it}

\makeatletter
\newcommand*{\defeq}{\mathrel{\rlap{%
                     \raisebox{0.3ex}{$\m@th\cdot$}}%
                     \raisebox{-0.3ex}{$\m@th\cdot$}}%
                     =}
\newcommand*{\eqdef}{=\mathrel{\rlap{%
                     \raisebox{0.3ex}{$\m@th\cdot$}}%
                     \raisebox{-0.3ex}{$\m@th\cdot$}}%
                     }
\makeatother

\usepackage{mathtools}

\usepackage{hhline}

\usepackage{stmaryrd}

\usepackage{enumitem}

\definecolor{light-gray}{gray}{.85}
\newsavebox{\songboxbox}

\usepackage[most]{tcolorbox}
\newtcbox{\MyBox}[1][red]{on line, size=tight, boxsep=1pt, colframe=#1!50!black, colback=#1!10!white}

\usepackage[framemethod=TikZ]{mdframed}
\mdfsetup{%
    outerlinewidth=1pt,
    innertopmargin=6pt,
    innerbottommargin=6pt,
    skipabove=10pt,
    skipbelow=10pt,
    backgroundcolor=black!10,
    roundcorner=20pt
}

\DeclareFontFamily{U}{matha}{\hyphenchar\font45}
\DeclareFontShape{U}{matha}{m}{n}{
      <5> <6> <7> <8> <9> <10> gen * matha
      <10.95> matha10 <12> <14.4> <17.28> <20.74> <24.88> matha12
      }{}
\makeatletter
\newcommand{\blandor}[1]{\mathbin{\@blandor{#1}}}
\newcommand{\@blandor}[1]{\mathchoice
  {\@@blandor{#1}{\tf@size}}
  {\@@blandor{#1}{\tf@size}}
  {\@@blandor{#1}{\sf@size}}
  {\@@blandor{#1}{\ssf@size}}
}
\newcommand{\@@blandor}[2]{%
    \raisebox{.1ex}{\rotatebox[origin=c]{#1}{%
      \fontsize{#2}{#2}\usefont{U}{matha}{m}{n}\symbol{\string"CE}}}%
}
\makeatother

\usepackage{slashed} %Feynman
\newcommand{\cmmnt}[1]{\ignorespaces} %for inline commenting
\usepackage{environ}
\NewEnviron{eqn}{
\begin{align}\begin{split}
\BODY
\end{split}\end{align}
}

\usepackage{bbding}
\DeclareFontFamily{U}{MnSymbolC}{}
\DeclareSymbolFont{MnSyC}{U}{MnSymbolC}{m}{n}
\DeclareMathSymbol{\diamondplus}{\mathbin}{MnSyC}{"7C}
\DeclareMathSymbol{\diamonddot}{\mathbin}{MnSyC}{"7E}
\DeclareFontShape{U}{MnSymbolC}{m}{n}{
    <-6>  MnSymbolC5
   <6-7>  MnSymbolC6
   <7-8>  MnSymbolC7
   <8-9>  MnSymbolC8
   <9-10> MnSymbolC9
  <10-12> MnSymbolC10
  <12->   MnSymbolC12}{}

\begin{document}

%\preprint{}

\title{Discrete Gravity with Local Lorentz Invariance} %Title of paper



\author{Eugene Kur}
%\thanks{}
\affiliation{
Lawrence Livermore National Laboratory, Livermore, CA 94550
}

\author{Alexander S. Glasser}
%\thanks{}
\affiliation{
Princeton Plasma Physics Laboratory, Princeton University, Princeton, New Jersey 08543\\
Department of Astrophysical Sciences, Princeton University, Princeton, New Jersey 08544
}

%\date{\today}

\hyphenpenalty=1000
\begin{abstract}
A novel structure-preserving algorithm for general relativity in vacuum is derived from a lattice gauge theoretic discretization of the tetradic Palatini action. The resulting model of discrete gravity is demonstrated to preserve local Lorentz invariance and symplectic structure.
\end{abstract}
\hyphenpenalty=50


\maketitle %\maketitle must follow title, authors, abstract and \pacs


%%%%%%%%%%%%%%%%%%%%%%%%%%%%%%%%%%%%%%%%%%%%%%%%%%%
%%%%%%%%%%%%%%%%%%%%%%%%%%%%%%%%%%%%%%%%%%%%%%%%%%%
%%%%%%%%%%%%%%%%%%%%%%%%%%%%%%%%%%%%%%%%%%%%%%%%%%%

\section{Introduction}

Since at least the 1990s, structure-preserving algorithms \cite{hairer_geometric_2006} have flourished in computational physics, having found wide adoption in subfields as diverse as orbital mechanics  \cite{kinoshita_symplectic_1991,gladman_symplectic_1991,chambers_symplectic_2002,bravetti_numerical_2020}, geophysics \cite{li_structure-preserving_2012,liu_modified_2015} and plasma physics \cite{squire_geometric_2012,xiao_explicit_2015,he_hamiltonian_2015,crouseilles2015Hamiltonian,qin_canonical_2016,kraus_gempic:_2017,morrison_structure_2017,glasser_geometric_2020,glasser_gauge-compatible_2021}. Such algorithms are generally derived from a Lagrangian or Hamiltonian formalism and use discretizations that preserve the symplectic structure, topology, gauge symmetry, and conservation laws of their underlying physical systems. This preservation of mathematical structure can substantially improve the accuracy and fidelity of numerical simulations.

Structure-preserving discretizations of general relativity (GR) arguably have an even longer history. The most widely explored such approach was introduced in 1961: Regge calculus \cite{regge_general_1961} is a discrete variational approximation of GR that encodes spacetime data on a simplicial mesh. In four spacetime dimensions, Regge calculus elegantly approximates the Einstein-Hilbert action by a sum over areas $A_h$ and \emph{deficit angles} $\delta_h$, such that
\begin{eqn}
S_\text{Regge}=\sum\limits_hA_h\delta_h~\xrightarrow{A_h\rightarrow0}~S_\text{EH}=\frac{1}{2}\int\md^4x \sqrt{-g} R
\end{eqn}
in the continuum limit. Here, $h$ labels each 2-simplex (i.e. triangle) of the simplicial complex, and $\delta_h$ describes the failure of the 4-simplices adjoining $h$---i.e. ${\{\sigma^4~|~\sigma^4\supset h\}}$---to tesselate their embedding in flat $\mR^4$ spacetime \cite{misner_gravitation_1973}.

Since the 1970s, Regge calculus has not only been actively employed as the basis of many studies in quantum gravity (e.g. \cite{hawking_spacetime_1978,caselle_regge_1989,williams_regge_1992,immirzi_quantum_1997,loll_discrete_1998,ambjorn_nonperturbative_2000,gionti_discrete_2005,dittrich_lorentzian_2021}), but also as an algorithmic approach to classical numerical relativity (e.g.  \cite{collins_application_1972,collins_dynamics_1973,sorkin_time-evolution_1975,porter_new_1987,dubal_relativistic_1989,barrett_parallelizable_1997,gentle_regge_2002,khavari_regge_2009,gentle_cosmological_2013}). Despite its success as a numerical tool, however, most studies in numerical relativity continue to depend upon standard finite difference methods. Two reasons cited for this include the need to develop (i) a description of matter in Regge calculus, as well as (ii) a better understanding of its relationship to standard methods in numerical relativity \cite{gentle_brief_2002,barrett_tullio_2019}.

In particular, because the degrees of freedom of Regge calculus are quite distinct from those of continuum GR, it can be challenging to initialize a Regge calculus simulation with known GR initial conditions, or to test whether a particular simulation using Regge calculus recovers a known GR solution. Although various physical solutions have indeed been thoroughly and successfully benchmarked with Regge calculus \cite{gentle_apparent_1999,gentle_simplicial_1999}, it would seem that any given simulation generally requires a bespoke understanding of the map between discrete and continuum degrees of freedom.

It is also worth emphasizing that, despite Regge calculus being a variational method, it nonetheless forfeits---in its complete, nonperturbative formulation---the local gauge symmetry of GR \cite{loll_discrete_1998}. While local gauge symmetry is maintained in a Regge calculus description of flat spacetime---and even in a linearized Regge calculus of curved spacetimes \cite{dittrich_covariant_2010}---this structural feature of GR is at best only partially preserved overall.

In this paper, an alternative variational approach to simulating general relativity is developed that ameliorates some of these limitations. Our effort employs familiar tools of lattice gauge theory \cite{kogut_introduction_1979} to construct a structure-preserving discretization of the tetradic Palatini action \cite{palatini_deduzione_1919}. Using a Poincar\'e group-valued connection derived from Cartan geometry, we describe a novel variational algorithm for numerical relativity that exactly preserves Lorentz gauge symmetry. We further show the algorithm is (multi)symplectic, with a symplectic structure analogous to continuum GR.

The approach we take is closely related to Poincar\'e gauge theoretic studies of lattice quantum gravity by Menotti, et al. \cite{menotti_poincare_1987,menotti_gauge_1987}. To our knowledge, however, the classical physics of these methods, including their equations of motion, for example, have not previously been explored, nor have they been extended to define an algorithm for numerical relativity. Moreover, our construction is general to simplicial and cubical discretizations of spacetime, and we develop a streamlined construction of the aforementioned Poincar\'e connection.

The remainder of this paper is organized as follows: Section~\ref{reviewPalatini} briefly reviews the tetradic Palatini action and its origins in Cartan geometry; Section~\ref{DiscreteActionSection} derives a discretization of this action in a manner that preserves Lorentz gauge invariance; Section~\ref{EOMSection} derives the discrete, classical equations of motion that comprise the algorithm; and Section~\ref{SymplecticSection} describes its symplectic structure. Finally, Section~\ref{ConclusionSection} summarizes and concludes.

\section{The Tetradic Palatini Action in Continuous Spacetime\label{reviewPalatini}}

Let us first review the tetradic Palatini action in the continuum. We consider a four-dimensional Lorentzian spacetime with connection, denoted ${(M,g,\Gamma)}$, and employ the following index conventions:
\begin{enumerate}[label=(\roman*)]
\setlength\itemsep{2pt}
\item spacetime coordinate indices $\{\mu,\nu,\dots\}$ are raised and lowered by ${g_{\mu\nu}}$, the metric on $M$;
\item internal Lorentz indices $\{A,B,\dots\}$ are raised and lowered by ${\eta_{AB}}$, the Minkowski metric; and
\item any other indices $\{a,b,\dots\}$ will be specified as needed.
\end{enumerate}
In a coordinate basis ${\{\partial_\mu\}}$, the affine connection $\Gamma$ has components ${\Gamma^\sigma_{\mu\nu}=\md x^\sigma(\nabla_{\partial_\mu}\partial_\nu)}$.

Up to local Lorentz gauge, the metric $g$ uniquely determines a tetrad field $e$ on $M$, a vector-valued 1-form with components ${e^A=e^A_\mu\md x^\mu\in\Gamma(T^*M)}$ defined to satisfy
\begin{eqn}
g_{\mu\nu}=e^A_\mu\eta_{AB}e^B_\nu.
\label{g_equal_e_eta_e}
\end{eqn}
Since $g_{\mu\nu}$ is non-degenerate, ${e^A_\mu(p)}$ defines $\forall$ ${p\in M}$ an isomorphism between the tangent space $T_pM$ and the `internal Lorentz space' at $p$. As a result, any vector field ${X\in\Gamma(TM)}$ can be equally well described in terms of the Lorentz frame ${\{\partial_A=e^\mu_A\partial_\mu\}}$ such that ${X=X^A\partial_A=X^\mu\partial_\mu}$. (Here, ${e^\mu_Ae^A_\nu=\delta^\mu_\nu}$ defines a matrix inverse.) In general, ${\{\partial_A\}}$ is a non-coordinate basis (since the commutator ${[\partial_A,\partial_B]}$ need not vanish) and is dual to ${\{e^A\}}$.

Parallel transport may be defined in the Lorentz frame by the 1-form spin connection $\omega$, with components ${\omega^A_{~B}=\omega^A_{\mu B}\md x^\mu\in\Gamma(T^*M)}$, such that
\begin{eqn}
\nabla_\mu X^A&=\partial_\mu X^A+\omega^A_{\mu B}X^B.
\end{eqn}
Since the Lorentz frame arises, ultimately, as a change of basis, the spin connection components have a definite relation to ${\Gamma^\sigma_{\mu\nu}}$. In particular, ${\omega^A_{\mu B}=e^A(\nabla_{\partial_\mu}\partial_B)}$, which can be more suggestively expanded as
\begin{eqn}
\nabla_\mu e^A_\nu=\partial_\mu e^A_\nu+\omega^A_{\mu B}e^B_\nu-\Gamma^\sigma_{\mu\nu}e^A_\sigma=0.
\label{e_compatibility}
\end{eqn}

The name `Lorentz frame' can be justified by requiring $\eta_{AB}$ to be invariant under parallel transport---
\begin{eqn}
\hspace{-15pt}0=\nabla_\mu\eta_{AB}&=\partial_\mu\eta_{AB}-\omega^C_{\mu A}\eta_{CB}-\omega^C_{\mu B}\eta_{AC}\\
&=-(\omega_{\mu BA}+\omega_{\mu AB}).
\label{eta_compatibility}
\end{eqn}
Due to its resulting antisymmetry, $\omega$ is defined by this condition as an $\mso$-valued 1-form. Studying Eqs.~(\ref{e_compatibility}) and (\ref{eta_compatibility}), we also see that the metric compatibility of ${\Gamma}$ follows immediately from this $\eta$-compatibility of $\omega$.

In Einstein GR, $\Gamma$ is assumed to be not only metric-compatible but torsion-free, such that ${\Gamma^\sigma_{[\mu\nu]}=0}$. The resulting Levi-Civita connection $\Gamma_\text{LC}$ is then uniquely determined by $g$. Thus, as a further consequence of vanishing torsion, by Eqs.~(\ref{g_equal_e_eta_e}) and (\ref{e_compatibility}) the metric $g$ also uniquely determines---up to Lorentz gauge---the ${\mR^4}$ and $\mso$-valued 1-forms $e$ and $\omega$, respectively, on $M$.

Conversely, the metric $g$ and the connection $\Gamma_\text{LC}$ can be uniquely recovered from the fields $e$ and $\omega$ on a torsion-free manifold. (More precisely, $g$ can be canonically recovered up to an overall constant factor.) There is an equivalence, therefore, between a Lorentzian manifold ${(M,g,\Gamma_\text{LC})}$ and its torsion-free \emph{Cartan geometric} counterpart ${(M,e,\omega)}$ \cite{sharpe_differential_1997}. Let us describe the origin of this nomenclature.

The 1-forms $e$ and $\omega$ are more economically regarded as components of the \emph{Cartan connection} ${A=A_\mu\md x^\mu}$ on $M$, defined by
\begin{eqn}
A = \left[\begin{matrix}\omega & e\\\v{0} & 0\end{matrix}\right]\in\Gamma(\mmp\otimes T^*M).
\end{eqn}
Here, ${\mmp=\mso\ltimes\mR^4\subset\mathfrak{gl}_5(\mR)}$ denotes the Lie algebra of the Poincar\'e group, so that
$A$ is a $\mmp$-valued 1-form on $M$.\footnote{A $\mmp$-valued Cartan connection $A$ is formally defined on an ${SO(3,1)}$-principal bundle $P$ over $M$ such that ${A:T_pP\rightarrow\mmp}$ is an isomorphism $\forall$ ${p\in P}$. The pair ${(P,A)}$ defines a \emph{Cartan geometry} \cite{sharpe_differential_1997}. In physical applications, however, $A$ is conventionally defined by its pullback to $M$ and its overlying bundle is elided.} Following the previous discussion, a torsion-free Lorentzian manifold can be equivalently defined by its metric $g$ or $\mmp$-valued Cartan connection $A$, and solving for the dynamics of $A$ similarly determines dynamics for $g$. In what follows, we therefore regard the tetradic Palatini action as a dynamical theory of the Cartan connection $A$ on $M$.

To that end, we first recall the curvature 2-form of the Cartan connection, defined as
\begin{eqn}
F=\md A+A\wedge A=\left[\hspace{2pt}\begin{matrix}[c|c]\md\omega+\omega\wedge\omega & \mD e\\\hline\v{0}&0\end{matrix}\hspace{2pt}\right]=\left[\begin{matrix}R & T\\\v{0}&0\end{matrix}\right],
\end{eqn}
where ${\mD e=\md e+\omega\wedge e}$ denotes the exterior covariant derivative of $e$. In components,
\begin{eqn}
R^A_{~B\mu\nu}&=\partial_\mu\omega^A_{\nu B}-\partial_\nu\omega^A_{\mu B}+\omega^A_{\mu C}\omega^C_{\nu B}-\omega^A_{\nu C}\omega^C_{\mu B}
\label{eq:define_F}
\end{eqn}
denotes the \emph{Lorentz curvature} ${R^A_{~B}\in\Gamma(\wedge^2T^*M)}$, while
\begin{eqn}
T^A_{\mu\nu}=(\mD e^A)_{\mu\nu}=\partial_\mu e^A_\nu-\partial_\nu e^A_\mu+\omega^A_{\mu B}e^B_\nu-\omega^A_{\nu B}e^B_\mu
\end{eqn}
denotes the \emph{torsion} ${T^A\in\Gamma(\wedge^2T^*M)}$. As previously noted, torsion is assumed to vanish, ${T^A=0}$, a priori in Einstein GR. In the tetradic Palatini theory, however, torsion does not vanish by assumption, but rather as a dynamical consequence of the action varied in vacuum, as we presently demonstrate.

The 4-form Lagrangian ${\mL_\text{Pal}\in\Gamma(\wedge^4T^*M)}$ of the tetradic Palatini action ${\mS_\text{Pal}=\int_M\hspace{-2pt}\mL_\text{Pal}}$ is  defined in terms of the tetrad $e$ and spin connection $\omega$ by \cite{palatini_deduzione_1919}
\begin{eqn}
\mL_\text{Pal}&=\epsilon_{ABCD}\left(e^A\wedge e^B\wedge R^{CD}\right).
\label{PalatiniLagrangian}
\end{eqn}
It should be noted that Eq.~(\ref{PalatiniLagrangian}) is often called the \emph{Einstein-Cartan-Sciama-Kibble} (ECSK) action \cite{hehl_general_1976,hehl_four_1980}. However, because ECSK theory prioritizes the role of torsion in gravity, whereas we will pursue only the torsionless vacuum equations of Einstein-Cartan gravity, we prefer the nomenclature \emph{tetradic Palatini action}. We note that the Lorentz invariance of $\mL_\text{Pal}$ follows directly from the $SO(3,1)$ invariance of the Levi-Civita symbol ${\epsilon_{ABCD}}$, that is,
\begin{eqn}
(\epsilon_{ABCD})'=\epsilon_{EFGH}\Lambda^E_{~A}\Lambda^F_{~B}\Lambda^G_{~C}\Lambda^H_{~D}=\epsilon_{ABCD}\ms{det}[\Lambda]
\end{eqn}
where ${\ms{det}[\Lambda]=1}$.

Unlike the fields ${(g,\Gamma_\text{LC})}$ of the Einstein-Hilbert action, $(e,\omega)$ are taken to be independent in Eq.~(\ref{PalatiniLagrangian}), and varied accordingly. The variation of each field yields the respective equations of motion \cite{menotti_lectures_2017}
\begin{eqn}
(\delta e):\hspace{20pt}&0=\epsilon_{ABCD}e^B\wedge R^{CD}\\
(\delta\omega):\hspace{20pt}&0=\epsilon_{ABCD}\mD(e^A\wedge e^B).
\label{PalatiniEOM}
\end{eqn}
Here, we note that ${\mD(e^A\wedge e^B)=\mD e^A\wedge e^B-e^A\wedge\mD e^B}$.
Taking ${R^{CD}=\frac{1}{2}R^{CD}_{~GH}e^G\wedge e^H}$ and ${T^A=\frac{1}{2}T^A_{GH}e^G\wedge e^H}$, (well-defined expansions for $e^A$ nondegenerate), it is readily established that the former relation ${(\delta e)}$ of Eq.~(\ref{PalatiniEOM}) yields Einstein's vacuum field equations while ${(\delta\omega)}$ yields a zero torsion condition. In particular, since ${\epsilon_{ABCD}e^A\wedge e^G\wedge e^H\wedge e^I=\delta^{GHI}_{BCD}e\ms{vol}}$ for a volume form $\ms{vol}$ and ${e=\ms{det}[e^A_\mu]}$, the wedge product ${(\delta e)\wedge e^I}$ gives
\begin{eqn}
0&=-\frac{1}{2}R^{CD}_{~~~GH}\delta^{GHI}_{ACD}=2R^{HI}_{~~~HA}-R^{GH}_{~~~GH}\delta^I_A,
\end{eqn}
which, using Eq.~(\ref{e_compatibility}), can be demonstrated equivalent to Einstein's vacuum equations, ${0=2R_{\mu\nu}-Rg_{\mu\nu}}$. Likewise, ${(\delta\omega)\wedge e^I}$ gives
\begin{eqn}
0=-\delta^{GHI}_{ACD}T^A_{GH}&=2\left(T^A_{CA}\delta^I_D+T^A_{AD}\delta^I_C-T^I_{CD}\right).
\label{torsionComp}
\end{eqn}
Tracing over Eq.~(\ref{torsionComp}) with $\delta^D_I$ in four dimensions leaves ${0=T^A_{CA}}$. By Eq.~(\ref{torsionComp}), therefore, ${T^I_{CD}=0}$ in all components, and torsion vanishes as desired.

Thus, despite making fewer initial assumptions, the tetradic Palatini action nevertheless recovers the equations of motion of GR in vacuum; the dynamics of the Cartan connection indeed recover those of GR.

Before concluding our discussion of continuous spacetime, the following will be useful for the next section, which discretizes $\mS_\text{Pal}$. Evaluated on a 4-tuple of vector fields---${\v{X}=(X_1,X_2,X_3,X_4)}$,  ${X_a\in\Gamma(TM)}$---the 4-form $\mL_\text{Pal}$ of Eq.~(\ref{PalatiniLagrangian}) yields
\begin{eqn}
\mL_\text{Pal}(\v{X})&=\frac{1}{2}\epsilon_{ABCD}(e^A_\mu e^B_\nu R^{CD}_{\sigma\tau})\epsilon^{abcd}X_a^\mu X_b^\nu X_c^\sigma X_d^\tau\\
&=\frac{1}{2}\epsilon_{ABCD}(e^A_\mu e^B_\nu R^{CD}_{\sigma\tau})\epsilon^{\mu\nu\sigma\tau}\ms{det}[\v{X}],
\label{continuousLagrangian}
\end{eqn}
where the function $\ms{det}[\v{X}]$ is the matrix determinant of the 4-tuple, expressed in the coordinate basis induced by ${\{x^\mu\}}$ and evaluated pointwise over $M$.

\section{The Discrete Action\label{DiscreteActionSection}}

We now discretize the tetradic Palatini action of the previous section by methodically mapping its continuum degrees of freedom to their discrete counterparts on a lattice. Our formalism will be general to orientable simplicial and cubical discretizations and we take care to preserve the theory's Lorentz invariance.

To proceed, we first choose a coordinate chart on the continuum spacetime manifold $M$, and construct a lattice (simplicial or cubical) on its coordinate space in $\mR^4$.  As such, the lattice inherits the Euclidean geometry of the coordinate space (such as straight edges and flat faces), but this `lattice geometry' will play no role in our description of spacetime. Topological features of $M$ must be retained in the construction of the lattice, including via possible identifications of its edges or faces. In such a case, the lattice should be regarded as only locally embedded in coordinate space while being globally homeomorphic to the target spacetime manifold. Such a construction is standard in the triangulation of manifolds (see e.g. \cite{nakahara2018geometry}).

To establish notation for lattice degrees of freedom, we denote the set of lattice $k$-cells by ${\Sigma^k=\{\sigma^k\}}$, such that an arbitrary oriented $k$-cell will be denoted $\sigma^k$, or will otherwise be specified by an ordered label of its vertices. ${\sigma_{ij}\in\Sigma^1}$, for example, denotes an edge oriented from vertex ${\sigma_i\in\Sigma^0}$ to vertex ${\sigma_j\in\Sigma^0}$. We define ${N_i(\sigma^k)=\{j\neq i~|~\sigma_{ij}\subset\sigma^k\}}$ as the set of labels of neighboring vertices in the cell $\sigma^k$ that share an edge with basepoint $\sigma_i$. In both simplicial and cubical discretizations in four dimensions, for example, ${\#N_i(\sigma^4)=4}$ if ${\sigma_i\subset\sigma^4}$ and 0 otherwise. We denote the permutation set of these neighboring vertex labels as ${\Pi_i(\sigma^4)=S[N_i(\sigma^4)]}$.

As described in Section~\ref{reviewPalatini}, the geometric information of $M$ is encoded in its Cartan connection---the fields $e^A_\mu(x)$ and $\omega^A_{\mu B}(x)$---which may be regarded as defined on coordinate space. A natural first (provisional) approximation of the tetradic Palatini action, as it appears in Eq.~(\ref{continuousLagrangian}), therefore follows by mapping these fields to the lattice, such that
\begin{eqn}
    \mS_\text{Pal} &= \sum_{\sigma^4\in\Sigma^4}\int_{\sigma^4}\mL_\text{Pal}(x)\md^4x\\
    &\approx \sum_{\substack{\sigma^4\in\Sigma^4\\\sigma_i\in\sigma^4}} \frac{(-1)^{|\pi|}V_f}{2n_v}\epsilon^{\mu\nu\sigma\tau}\epsilon_{ABCD}\Big(\hspace{-2pt}e^A_\mu e^B_\nu R^{CD}_{\sigma\tau}\hspace{-2pt}\Big)\Big|_{\sigma_i}\hspace{-2pt}\ms{det}[\mathbf{V}_{\sigma_i}]
    \label{eq:action_first_approx}
\end{eqn}
where ${\mL_\text{Pal}(x)}$ denotes $\mL_\text{Pal}$ on coordinate space. The second line approximates the integral over ${\sigma^4}$ by averaging the value of its integrand, as expressed in Eq.~(\ref{continuousLagrangian}), at each of its vertices. More specifically:
\begin{itemize}
    \item $n_v$ denotes the number of vertices in $\sigma^4$ over which the integrand is averaged. ${n_v=5}$ on a simplicial lattice and ${n_v=16}$ on a cubical lattice.
    \item ${\mathbf{V}_{\sigma_i}=(V_{i1},V_{i2},V_{i3},V_{i4})}$ is a 4-tuple of `edge vectors' emanating from vertex $\sigma_i$. These point in the directions of neighboring vertices in $\sigma^4$, with magnitudes set by the edges' coordinate lengths.
    \item $(-1)^{|\pi|}$ accounts for the relative orientation between the 4-tuple $\mathbf{V}_{\sigma_i}$ and the cell $\sigma^4$, whose orientation is inherited from $M$. This factor is expressed in terms of a permutation $\pi$ to be defined more concretely below.
    \item The volume factor $V_f$ corrects for the fact that ${\ms{det}[\mathbf{V}_{\sigma_i}]}$ implicitly evaluates ${\mL_\text{Pal}}$ not on $\sigma^4$, but on a (hyper-) parallelapiped specified by $\mathbf{V}_{\sigma_i}$. On a cubical lattice, these volumes coincide and ${V_f=1}$, but on a simplicial lattice, ${\ms{det}[\mathbf{V}_{\sigma_i}]}$ overcounts the volume of $\sigma^4$ by the ratio of a normalized hypercube to one of its corners, such that ${V_f}=1/4!$.
\end{itemize}
This approximation of $\int_{\sigma^4}$ by the average of vertex evaluations is, in effect, a second order accurate multi-dimensional trapezoid rule (see \cite{haber1970numerical} and references therein).

Eq.~(\ref{eq:action_first_approx}) instructively approximates the continuum action, but it is insufficient to determine dynamics for a discrete theory. In particular, Eq.~(\ref{eq:action_first_approx}) discretely samples degrees of freedom that are manifestly defined in the continuum---e.g. ${\partial\omega}$. (If this continuous derivative were avoided by regarding $R$ itself as a Lie-algebra-valued degree of freedom, rather than $\omega$, the resulting action would not recover the equations of motion of Einstein GR.) To reformulate Eq.~(\ref{eq:action_first_approx}) with bona fide discrete degrees of freedom, we now proceed hewing more closely to the underlying differential geometry of the tetradic Palatini action.

In particular, the study of structure-preserving discretizations (such as discrete exterior calculus (DEC) \cite{desbrun_discrete_2005} and finite element exterior calculus (FEEC) \cite{arnold_finite_2006,arnold_finite_2010}) has demonstrated the importance of preserving the degrees of discrete differential forms. Therefore, rather than sampling continuum fields at vertices, as we do in Eq.~(\ref{eq:action_first_approx}), we will instead map 1-forms to data associated with edges, and 2-forms to data associated with faces.

However, an additional challenge we must overcome is the gauge-dependent character of the fields we are modeling, which thwarts conventional approaches such as DEC and FEEC. The preservation of Lorentz invariance in our theory will require that we express discrete fields in a definite (if arbitrary) Lorentz gauge, which is associated in a continuum theory to each point of spacetime, and in a discrete theory to each vertex. This pointwise gauge selection is in tension with the desire to characterize differential forms over edges and faces of finite extent. For example, a scalar-valued 1-form is conventionally approximated on an edge by its integral over that edge. Here, such an integral involves a continuum of different gauge choices in spacetime that prevent the simple summation of fields defined at disparate points.

A resolution to this tension is naturally found in the holonomy of a connection. The path-ordered integral of a 1-form connection produces the means to parallel transport between different gauge choices. It is an object that can be naturally associated with an edge, and which by construction accounts for a difference in gauge between two vertices. In this sense, the Cartan connection---which retains the geometric data of a Lorentzian manifold--- provides a natural approach to a structure-preserving discretization of GR.

To map the Cartan connection $A$ on $M$ to holonomies on the discrete lattice in coordinate space, we associate to each edge $\sigma_{ij}$ the following path-ordered integral:
\begin{eqn}
U_{ij}=\mmP\left\{\exp\hspace{-3pt}\int_{\sigma_{ij}}\hspace{-3pt}A\right\}=\mmP\left\{\exp\hspace{-3pt}\int_{\sigma_{ij}}\hspace{-3pt}\left[\begin{matrix}\omega&e\\\v{0}&0\end{matrix}\right]\right\}=\left[\begin{matrix}\Lambda_{ij} & \ell_{ij}\vspace{2pt}\\\v{0}&1\end{matrix}\right].
\label{path_int_holonomy}
\end{eqn}
This is a standard construction of lattice gauge theory \cite{kogut_introduction_1979}. $U_{ij}$ constitutes the Poincar\'e group-valued holonomy associated with edge $\sigma_{ij}$, expressed in the representation ${SO(3,1)\ltimes\mR^4\subset GL_5(\mR)}$ and characterized by Lorentz and translation group elements, ${\Lambda_{ij}\in SO(3,1)}$ and ${\ell_{ij}\in\mR^4}$, respectively.

We denote the ${(A,B)^\text{th}}$ component of the Lorentz connection along edge $\sigma_{ij}$ by $\Lambda^A_{ijB}$, and the ${A^\text{th}}$ component of the corresponding translation connection by $\ell^A_{ij}$. We also adopt a notation for a Lorentz holonomy with an arbitrary number of edges. In particular,  for the holonomy comprised of ${(n-1)}$ connections between the vertices ${\sigma_{i_1},\dots,\sigma_{i_n}}$, we write
\begin{eqn}
\Lambda^A_{i_1\cdots i_nB}&=\big(\Lambda_{i_1i_2}\Lambda_{i_2i_3}\cdots\Lambda_{i_{n-1}i_n}\big)^A_{~B}\\
&=\Lambda^A_{i_1i_2C}\Lambda^{C}_{i_2i_3D}\cdots\Lambda^E_{i_{n-1}i_nB}
\label{holonomyNotation}
\end{eqn}
where intermediate Lorentz indices ${\{C,D,\dots,E\}}$ are all contracted. Here, we have implicitly defined the holonomy to act from the right, and note that the matrix multiplication of holonomies effects the concatenation of path-ordered integrals, as defined in Eq.(\ref{path_int_holonomy}).

$U_{ij}$ is seen to `mediate' between Lorentz gauges at $\sigma_i$ and $\sigma_j$, as desired. In particular, given an arbitrary Lorentz gauge transformation defined at each vertex, say
\begin{eqn}
\Big\{g_i=g(\sigma_i)\in SO(3,1)~\forall~\sigma_i\in\Sigma^0\Big\},
\end{eqn}
the gauge transformation of $U_{ij}$ readily follows from Eq.~(\ref{path_int_holonomy}), such that
\begin{eqn}
U_{ij}'&=\left[\begin{matrix}g_i&\v{0}\vspace{2pt}\\\v{0}&1\end{matrix}\right]^{-1}U_{ij}\left[\begin{matrix}g_j&\v{0}\vspace{2pt}\\\v{0}&1\end{matrix}\right]=\left[\begin{matrix}[c|c]g_i^{-1}\Lambda_{ij}g_j & g_i^{-1}\ell_{ij}\\\hline\v{0}&1\end{matrix}\right].
\label{lorentzGaugeTransf}
\end{eqn}
From this calculation, we note that ${\ell_{ij}}$  can be regarded as if `based at' $\sigma_i$. By examining Eq.~(\ref{path_int_holonomy}), the holonomy $U_{ji}$ is also readily calculated to be ${U_{ji}=U_{ij}^{-1}}$. In particular,
\begin{eqn}
\Lambda_{ji}=(\Lambda_{ij})^{-1}~~\text{and}~~ \ell_{ji}=-\Lambda_{ji}\ell_{ij}.
\label{inverseRepresentation}
\end{eqn}

We must consider holonomies on closed paths as well. For example, given a loop ${(\partial\sigma^2)_i=\sigma_i\sigma_j\cdots\sigma_k\sigma_i}$ around a single face $\sigma^2$ with basepoint $\sigma_i$, we define
\begin{eqn}
\left[\begin{matrix}\Omega_{ijk}&\Theta_{ijk}\\0&1\end{matrix}\right]=\mmP\left\{\exp\hspace{-2pt}\oint_{(\partial\sigma^2)_i}\hspace{-4pt}A\right\}.
\end{eqn}
Note, when a holonomy is comprised of the connections along the edges of a single face ${\sigma^2\in\Sigma^2}$ (i.e. when it is a ``minimal" nontrivial loop), we use the symbol $\Omega$ for its Lorentz holonomy rather than $\Lambda$ as in Eq.~(\ref{holonomyNotation}), and we suppress some of its indices. This notation is general to the simplicial and cubical setting, such that, for example,
\begin{alignat}{3}
\hspace{-7pt}\text{Simplicial:}&~~&\Omega^{AB}_{ijk}=&&&\big(\Lambda_{ij}\Lambda_{jk}\Lambda_{ki}\big)^A_{~C}\eta^{CB}\nonumber\\
\hspace{-7pt}\text{Cubical:}&~~&\Omega^{AB}_{ijk}=&&&\big(\Lambda_{ij}\Lambda_{ji'}\Lambda_{i'k}\Lambda_{ki}\big)^A_{~C}\eta^{CB}.
\label{curvatureDefinitions}
\end{alignat}
Here, $i'$ labels the vertex diagonal to $i$ on the appropriate face of a cubical lattice; in a more typical notation, ${(i,j,i',k)=(\v{n},\v{n}+\hat{a},\v{n}+\hat{a}+\hat{b},\v{n}+\hat{b})}$. $\Omega$ thereby characterizes Lorentz curvature over a face $\sigma^2$, while $\Theta$ characterizes the corresponding torsion.

To see how these holonomies can be substituted for the fields of Eq.~(\ref{eq:action_first_approx}), let us examine their continuous limit. We Taylor expand around $\sigma_i$ to find \cite{kogut_introduction_1979}
\begin{eqn}
\Lambda_{ij}&\approx\mOne+\omega_{ij}(\sigma_i)\Delta+\omega_{ij}(\sigma_i)^2\frac{\Delta^2}{2}+\mO(\Delta^3)\\
\ell_{ij}&\approx e_{ij}(\sigma_i)\Delta+\omega_{ij}(\sigma_i)e_{ij}(\sigma_i)\frac{\Delta^2}{2}+\mO(\Delta^3)\\
\Omega_{ijk}-\Omega_{ikj}&\approx2A_fR_{ijk}(\sigma_i)\Delta^2+\mO(\Delta^3).
\label{contExpansions}
\end{eqn}
Here, ${\omega_{ij}(\sigma_i)}$ denotes the component of the continuum Lorentz connection along the lattice edge ${\sigma_{ij}}$, evaluated at $\sigma_i$. ${e_{ij}(\sigma_i)}$ is defined analogously. $R_{ijk}(\sigma_i)$ denotes the component of the continuum Lorentz curvature at $\sigma_i$ corresponding to edge vectors ${\sigma_{ij}}$ and ${\sigma_{ik}}$. $\Delta$ denotes the length of $\sigma_{ij}$ and $\sigma_{ik}$ in coordinate space (in this expansion we assume these to be equal for simplicity, though they need not be in general), and $\omega$, $e$, and $R$ are implicitly expressed in the corresponding coordinate basis. The area factor $A_f$ is analogous to $V_f$ in Eq.~(\ref{eq:action_first_approx})---it corrects for the implicit overcounting of area in the simplicial setting on the parallelogram formed by $\sigma_{ij}$ and $\sigma_{ik}$. In particular, ${A_f=1}$ on a cubical lattice and ${A_f=1/2}$ on a simplicial lattice. It is further worth noting that the difference ${\Omega-\Omega^{-1}}$ in Eq.~(\ref{contExpansions}) is, in fact, $\mso$-valued, since ${(\Omega-\Omega^{-1})^T\eta=\eta(\Omega^{-1}-\Omega)}$ $\forall$ ${\Omega\in SO(3,1)}$.

With the expansions of Eq.~(\ref{contExpansions}) in mind, it is now straightforward to reconstruct a discrete tetradic Palatini action using edge holonomies, such that Eq.~\eqref{eq:action_first_approx} is recovered to least order in the continuum limit. In particular, we define the following  action summed over lattice hypercells ${\{\sigma^4\}=\Sigma^4}$:
\begin{eqn}
S&=\smashoperator{\sum\limits_{\sigma^4\in\Sigma^4}}L(\sigma^4)\\
L(\sigma^4)&=\smashoperator{\sum\limits_{\substack{\sigma_i\subset\sigma^4\\\pi\in \Pi_i(\sigma^4)}}}\frac{(-1)^{\abs{\pi}}}{2\rho_f n_v}\epsilon_{ABCD}\left(\ell^A_{i\pi(1)}\ell^B_{i\pi(2)}\Omega^{CD}_{i\pi(3)\pi(4)}\right).
\label{discreteLagrangian}
\end{eqn}
With factors $V_f$ and $A_f$ as defined above, the quantity ${\rho_f=V_f/A_f}$ satisfies ${\rho_f=1}$ (${\rho_f=12}$) for cubical (simplicial) lattices. Note, we need not explicitly antisymmetrize $\Omega$ and $\Omega^{-1}$ because the Levi-Civita symbol does this for us. This cancels the factor of $2$ appearing in Eq.~(\ref{contExpansions}). The sum over permutations $\pi$ replaces $\epsilon^{\mu\nu\sigma\tau}$ in Eq.~\eqref{eq:action_first_approx} and the corresponding summation of spacetime indices. In particular, as first introduced in Eq.~(\ref{eq:action_first_approx}), ${\pi\in\Pi_i(\sigma^4)}$ is now explicitly defined as a permutation of vertices neighboring $\sigma_i$ in $\sigma^4$. We define the parity $|\pi|$ to correct for any disagreement between the overall orientation of coordinate space and the orientation of edge vectors in $\sigma^4$, emanating from $\sigma_i$ and ordered by $\pi$.

It is worth emphasizing the following important features of this discrete action:
\begin{itemize}
    \item The discrete Lagrangian $L(\sigma^4)$ is locally Lorentz invariant. Under an arbitrary gauge transformation ${\{g_i\in SO(3,1)\}_{\sigma_i\in\Sigma^0}}$ using Eq.~(\ref{lorentzGaugeTransf}) we find
\begin{eqn}
\Big(&\epsilon_{ABCD}\ell^A_{i\pi(1)}\ell^B_{i\pi(2)}\Omega^{CD}_{i\pi(3)\pi(4)}\Big)'\\
&=\epsilon_{ABCD}\big(g_i^{-1}\ell_{i\pi(1)}\big)^A\big(g_i^{-1}\ell_{i\pi(2)}\big)^B\big(g_i^{-1}\Omega_{i\pi(3)\pi(4)}g_i\big)^{CD}\\
&=\epsilon_{ABCD}\ell^A_{i\pi(1)}\ell^B_{i\pi(2)}\Omega^{CD}_{i\pi(3)\pi(4)}.
\end{eqn}
The last equality above follows from the Lorentz group relation ${(g_i)^E_{~F}\eta^{FD}=(g_i^{-1})^D_{~F}\eta^{FE}}$ and the $SO(3,1)$-invariance of the Levi-Civita symbol.
\item The Poincar\'e holonomies ${U_{ij}=(\Lambda_{ij},\ell_{ij})}$ are not to be confused with the Poincar\'e symmetry group of Minkowski spacetime. There is a gauge symmetry transformation that acts on our Poincar\'e holonomies, but the gauge group is Lorentz, not Poincar\'e. Such `internal' or `vertical' (e.g. Lorentz) gauge groups are typical in Cartan geometries, despite their connections' `external' or `horizontal' (e.g. translation) components \cite{sharpe_differential_1997}. Even as the internal Lorentz gauge symmetry of our theory transforms the tetrad, it leaves spacetime geometry (i.e. the metric) completely unaffected---regardless of what (global) symmetries the geometry may or may not possess. By contrast, the Poincar\'e symmetry group of Minkowski spacetime is comprised of global transformations of the spacetime. It is a subgroup of the full diffeomorphism group that leaves the Minkwoski metric invariant. (I.e., the metric is invariant only if it happens to be Minkowski). This distinction means, in particular, that our use of Poincar\'e holonomies should not be taken to imply that we are describing Minkowski spacetime. Indeed, our theory is capable of describing any (discrete) spacetime. 
\end{itemize}

\section{The Discrete Equations of Motion\label{EOMSection}}

We now compute equations of motion (EOM) by varying the discrete action with respect to the connection. To compactify notation, when an element of the permutation $\pi$ appears in an index, it will hereafter be denoted only by a corresponding underlined number, for example, ${\underline{1}=\pi(1)}$. As usual in a first-order formalism, we assume $\Lambda_{ij}$ and $\ell_{ij}$ to be independent. Varying the action with respect to $\ell^A_{ij}$, and applying the expression for $U_{ji}$ from Eq.~(\ref{inverseRepresentation}) where appropriate, we find
\begin{eqn}
0=\frac{\partial S}{\ell^A_{ij}}=&\sum\limits_{\sigma^4\supset\sigma_{ij}}\Biggr[\sum\limits_{\substack{\pi\in \Pi_i(\sigma^4)\\\pi(1)=j}}\frac{(-1)^{\abs{\pi}}}{\rho_f n_v}\epsilon_{ABCD}\ell^B_{i\underline{2}}\Omega^{CD}_{i\underline{3}\underline{4}}\\
&-\sum\limits_{\substack{\pi\in \Pi_j(\sigma^4)\\\pi(1)=i}}\frac{(-1)^{\abs{\pi}}}{\rho_f n_v}\epsilon_{EBCD}\Lambda^E_{jiA}\ell^B_{j\underline{2}}\Omega^{CD}_{j\underline{3}\underline{4}}\Biggr].
\label{discreteEinsteinEquation}
\end{eqn}
%\begin{eqn}
%0=&\frac{\partial S}{\ell^A_{ij}}=\frac{1}{n_v}\sum\limits_{\sigma^4\supset\sigma_{ij}}\Biggr[~~\smashoperator{\sum\limits_{\pi\in \Pi_i(\sigma^4)}}\frac{(-1)^{\abs{\pi}}}{\rho_f}\epsilon_{ABCD}\delta_{j\underline{1}}\ell^B_{i\underline{2}}\Omega^{CD}_{i\underline{3}\underline{4}}\\
%&-\smashoperator{\sum\limits_{\theta\in \Pi_j(\sigma^4)}}\frac{(-1)^{|\theta|}}{\rho_f}\epsilon_{EBCD}(\Lambda_{ji})^{E}_{~A}\delta_{i\theta(1)}\ell^B_{j\theta(2)}\Omega^{CD}_{j\theta(3)\theta(4)}\Biggr]
%\label{discreteEinsteinEquation}
%\end{eqn}
%where $\delta_{ij}$ denotes the Kronecker delta symbol.
The first sum of Eq.~(\ref{discreteEinsteinEquation}) arises from terms with basepoint $i$ and the second from terms with basepoint ${j}$. We note that although frames are permuted at distinct basepoints in these two lines, their parities are understood to be induced by a global orientation and are therefore mutually consistent. Eq.~(\ref{discreteEinsteinEquation}) is counterpart to $(\delta e)$ of Eq.~(\ref{PalatiniEOM}), and constitutes a discrete reformulation of Einstein's vacuum equations. 

In particular, we may examine the continuum limit of Eq.~(\ref{discreteEinsteinEquation}) on a cubical lattice coordinatized by $\{x^\mu\}$ with regular lattice spacing $\Delta$. Expanding each degree of freedom near $\sigma_i$ as in Eq.~(\ref{contExpansions}) and taking edge vectors along coordinate directions (e.g. ${\sigma_{ij}\parallel\partial_\mu}$), we find at leading order $\mO(\Delta^3)$,
\begin{align*}
0=\epsilon^{\mu\nu\sigma\tau}\epsilon_{ABCD}e^B_\nu R^{CD}_{~~\sigma\tau}
\end{align*}
---mirroring ($\delta e$) of Eq.~(\ref{PalatiniEOM}).

We now derive the the Lorentz connection EOM, exercising caution to ensure that the variation of $\Lambda_{ij}$ is constrained to the $SO(3,1)$ manifold. In particular, ${\Lambda^T\eta\Lambda=\eta}$ implies ${(\Lambda^{-1}\delta\Lambda)^T\eta+\eta(\Lambda^{-1}\delta\Lambda)=0}$, so that ${\Lambda^{-1}\delta\Lambda\in\mathfrak{so}(3,1)}$ for a variation $\delta\Lambda$. We can impose this constraint by taking a variation that satisfies ${(\Lambda^{-1}\delta\Lambda)^{AB}=(\Lambda^{-1}\delta\Lambda)^{[AB]}}$, but is otherwise arbitrary.

To that end, we consider as an example the variation of ${\Omega^{AB}_{i\underline{3}\underline{4}}=\big(\Lambda_{i\underline{3}}\Lambda_{\underline{3}\underline{4}}\Lambda_{\underline{4}i}\big)^{AB}}$ in the simplicial setting with respect to the Lorentz connection on the edge $\sigma_{i\underline{3}}$:
\begin{align*}
\delta\Omega^{AB}_{i\underline{3}\underline{4}}&=\left(\Lambda^A_{i\underline{3} C}\Lambda^C_{\underline{3}iD}\right)\delta\Lambda^D_{i\underline{3}E}\Lambda^E_{\underline{3}\underline{4}F}\Lambda^{FB}_{\underline{4}i}\\
&=\Lambda^A_{i\underline{3} C}\Lambda^B_{i\underline{4}\underline{3}E}\left(\Lambda^{[C|}_{\underline{3}iD}\delta\Lambda^{D|E]}_{i\underline{3}}\right)\\
&=\Lambda^A_{i\underline{3} [C|}\Omega^B_{i\underline{43}F}\Lambda^F_{i\underline{3}|E]}\left(\Lambda^{[C|}_{\underline{3}iD}\delta\Lambda^{D|E]}_{i\underline{3}}\right).
\end{align*}
The term in parentheses on the first line is a conveniently chosen form of $\delta^A_D$---the Kronecker delta. The second line follows from the notation of Eq.~(\ref{holonomyNotation}), the identity ${(\Lambda^{-1})^{AB}=\Lambda^{BA}}$, and from asserting the antisymmetry of the variation ${(\Lambda^{-1}\delta\Lambda)^{[CE]}}$. The third line follows after inserting another Kronecker delta to form a closed loop holonomy. (In general, when $\Omega$ based at $\sigma_i$ is varied with respect to its Lorentz connection along $\sigma_{jk}$, the result can be expressed in terms of $\Omega$ or $\Omega^{-1}$ along with two antisymmetrized Lorentz transformations that effect a parallel transport from $\sigma_i$ to $\sigma_k$.)

To further facilitate the variation of the action, we introduce a couple concise notations. For brevity, we denote
\begin{align*}
a^{\abs{\pi}}=\frac{(-1)^{\abs{\pi}}}{\rho_f n_v}
\end{align*}
and also define
\begin{align*}
\mmP^{k\underline{12}}_{CD}=\epsilon_{ABCD}\ell_{k\underline{1}}^A\ell_{k\underline{2}}^B.
\end{align*}
$\mmP$ is antisymmetric both in its Lorentz indices and vertex permutation indices. It can be roughly regarded as a (non-idempotent) projection that annihilates any $\ell\in\ms{span}\{\ell_{k\underline{1}},\ell_{k\underline{2}}\}$.

Continuing in this way, we vary $S$ with respect to ${(\Lambda_{ji}\delta\Lambda_{ij})^{[MN]}}$ to find, in the simplicial case:
\begin{eqn}
\hspace{-5pt}0&=\frac{\partial S_{\text{simplicial}}}{(\Lambda_{ji}\delta\Lambda_{ij})^{[MN]}}\\
&=\sum\limits_{\sigma^4\supset{\sigma_{ij}}}\Biggr[\sum\limits_{\substack{\pi\in \Pi_i(\sigma^4)\\\pi(3)=j}}\hspace{-10pt}a^{\abs{\pi}}\Big(\mmP^{i\underline{12}}_{CD}\Omega^D_{i\underline{4}jE}\Big)\Lambda^E_{ij[M|}\Lambda^C_{ij|N]}\\
+&\sum\limits_{\substack{k\in\sigma^4\\k\neq i,j}}\sum\limits_{\substack{\pi\in \Pi_k(\sigma^4)\\\pi(3)=i\\\pi(4)=j}}\hspace{-10pt}a^{\abs{\pi}}\Big(\mmP^{k\underline{12}}_{CD}\Omega^D_{kijE}\Big)\Lambda^E_{kj[M|}\Lambda^C_{kj|N]}\\
+&\hspace{5pt}\sum\limits_{\substack{\pi\in \Pi_j(\sigma^4)\\\pi(4)=i}}\hspace{-10pt}a^{\abs{\pi}}\Big(\mmP^{j\underline{12}}_{CD}\Omega^D_{j\underline{3}iE}\Big)\delta^E_{[M|}\delta^C_{|N]}\Biggr].
\label{simplicialLorentzVariation}
\end{eqn}
The first line of Eq.~(\ref{simplicialLorentzVariation}) arises from terms with basepoint $i$, the middle line from terms with basepoint ${k\neq i,j}$ in $\sigma^4$ and the last from terms with basepoint ${j}$. The Lorentz EOM for a cubical discretization follows similarly:
\begin{eqn}
\hspace{-5pt}0&=\frac{\partial S_{\text{cubic}}}{(\Lambda_{ji}\delta\Lambda_{ij})^{[MN]}}\\
&=\sum\limits_{\sigma^4\supset{\sigma_{ij}}}\Biggr[\sum\limits_{\substack{\pi\in \Pi_i(\sigma^4)\\\pi(3)=j}}\hspace{-10pt}a^{|\pi|}\Big(\mmP^{i\underline{12}}_{CD}\Omega^D_{i\underline{4}jE}\Big)\Lambda^{E}_{ij[M|}\Lambda^C_{ij|N]}\\
&+\sum\limits_{\substack{k\in\sigma^4\\k\neq i,j}}\sum\limits_{\substack{\pi\in \Pi_k(\sigma^4)\\\pi(3)=i\\k'=j}}\hspace{-10pt}a^{|\pi|}\Big(\mmP^{k\underline{12}}_{CD}\Omega^D_{k\underline{4}iE}\Big)\Lambda^{E}_{kij[M|}\Lambda^C_{kij|N]}\\
&+\sum\limits_{\substack{k\in\sigma^4\\k\neq i,j}}\sum\limits_{\substack{\pi\in \Pi_k(\sigma^4)\\\pi(4)=j\\k'=i}}\hspace{-10pt}a^{|\pi|}\Big(\mmP^{k\underline{12}}_{CD}\Omega^D_{k\underline{3}jE}\Big)\Lambda^{E}_{kj[M|}\Lambda^C_{kj|N]}\\
&+\sum\limits_{\substack{\pi\in \Pi_j(\sigma^4)\\\pi(4)=i}}a^{\abs{\pi}}\Big(\mmP^{j\underline{12}}_{CD}\Omega^D_{j\underline{3}iE}\Big)\delta^E_{[M|}\delta^C_{|N]}\Biggr].
\label{cubicLorentzVariation}
\end{eqn}

Eqs.~(\ref{simplicialLorentzVariation}-\ref{cubicLorentzVariation}) enforce discrete zero-torsion conditions analogous to $(\delta\omega)$ of Eq.~(\ref{PalatiniEOM}). Again employing the ordering of Eq.~(\ref{contExpansions}) on a cubical lattice, it is readily computed that the least nontrivial contribution to Eq.~(\ref{cubicLorentzVariation}) is ${\mO(\Delta^3)}$ and arises from its second and third lines alone, with ${\Omega^D_{~E}=\delta^D_E}$. This leading order expression is given by
\begin{align*}
0&=\epsilon^{\mu\nu\alpha\beta}\epsilon_{ABNM}\Big[\partial_\nu\left(e^A_\alpha e^B_\beta\right)+\omega^A_{\nu I}e^I_\alpha e^B_\beta+\omega^B_{\nu J}e^A_\alpha e^J_\beta\Big],
\end{align*}
mirroring ($\delta\omega$) of Eq.~(\ref{PalatiniEOM}).

Eqs.~(\ref{discreteEinsteinEquation}-\ref{cubicLorentzVariation}) define the desired algorithm for vacuum numerical relativity. However, while these equations suffice to compute simulation steps in the \emph{bulk}, the evolution of \emph{boundary connections}---including connections along both spacelike and timelike boundaries---still requires some explanation. In particular, even if initial and boundary connections are known a priori, Eqs.~(\ref{discreteEinsteinEquation}-\ref{cubicLorentzVariation}) involve data from holonomies that generally extend outside of the boundary wall, and are therefore underspecified on the boundary.

The strategy we adopt \cite{stern_geometric_2009} to derive equations of motion for boundary connections, therefore, is to extend all spacelike and timelike boundary surfaces outward from the bulk, creating a narrow `double wall' of some fiducial thickness $\epsilon$ around the simulation domain. This double wall is then populated with cells of width $\epsilon$, such that connections between an inner wall vertex $\sigma_{i_\text{in}}$ and an outer wall vertex $\sigma_{i_\text{out}}$ will have ${\Lambda_{i_\text{in}i_\text{out}}\sim\mOne+\mO(\epsilon)}$ and ${\ell_{i_\text{in}i_\text{out}}\sim\mO(\epsilon)}$. The connections lying along the outer wall itself are chosen to copy the initial or boundary conditions of the inner wall. Then, equations of motion for the inner wall connections can be derived as usual from Eqs.~(\ref{discreteEinsteinEquation}-\ref{cubicLorentzVariation}), as they now behave as connections in the bulk. Finally, we take ${\epsilon\rightarrow0}$ in the resulting equations of motion for the (inner wall) boundary connections.

It is worth noting that not all boundary and initial conditions will satisfy the discrete equations of motion. Just as boundary constraints must be satisfied in the continuum theory, care must be taken to ensure that Eqs.~(\ref{discreteEinsteinEquation}-\ref{cubicLorentzVariation}) are satisfied on the initial surfaces of the discrete theory.

\section{Symplectic Structure of the Discrete Action\label{SymplecticSection}}
Variational integrators for field theories have a natural multisymplectic structure (see e.g. Refs.~\cite{marsden1998multisymplectic,gotay2006momentum} and references therein), generalizing the ordinary symplectic structure possessed by variational integrators in particle mechanics \cite{marsden2001discrete}. Here we review the proof that variational integrators are naturally (multi)symplectic, thereby confirming the multisymplectic structure of Eqs.~(\ref{discreteEinsteinEquation}-\ref{cubicLorentzVariation}).

In a variational integrator for particle mechanics, the action evaluated on a temporal cell ${[t_i,t_{i+1}]}$ provides a generating function for a canonical (symplectic) transformation across the cell \cite{marsden2001discrete}. Specifically, the discrete action $S_i$ for cell $i$ is a generating function for the canonical transformation $(q_i,p_i)\to (q_{i+1},p_{i+1})$, where $q_i$, $q_{i+1}$ are the particle coordinates at the left and right end of the cell respectively, $p_i=-\frac{\partial S_i}{\partial q_i}$, and $p_{i+1} = \frac{\partial S_i}{\partial q_{i+1}}$. The equations of motion (e.g. $\frac{\partial S_{i-1}}{\partial q_i}+\frac{\partial S_{i}}{\partial q_i}=0$) guarantee that the momentum at a point is identical whether using the left or right cell to define it. In this way the symplectic transformations inside the cells are glued consistently across cells to produce a global symplectic evolution.

In field theory, the situation is slightly different. A field $\phi$ has a multimomentum $\pi^\mu$ (one for each dimension of space-time) \cite{gotay1998momentum}, which in the case of a scalar field can be recast as a 3-form $\pi = \pi^\mu \mathrm{d}^3x_\mu=\frac{1}{3!}\epsilon_{\mu\alpha\beta\gamma}\pi^\mu\,\mathrm{d}x^\alpha\wedge\mathrm{d}x^\beta\wedge\mathrm{d}x^\gamma$. For any spacetime region $R$, we then have the boundary fields $\phi(\sigma),\pi(\sigma)$ living on $\Sigma=\partial R$, where $\pi(\sigma)=\pi\big|_\Sigma$ can be regarded as a pseudo-scalar field. The action evaluated over $R$, $S(R)$, is a generating function for a submanifold $\left\{\left(\phi(\sigma),\pi(\sigma)=\frac{\delta S(R)}{\delta \phi}\right)\right\}$ in this ``boundary phase space''. One may regard $S(R)$ (imprecisely) as a generating function for a canonical transformation between any two parts of the boundary. In the case when the boundary of $R$ consists of two disconnected pieces corresponding to two different times, $S(R)$ is the generating function for a canonical transformation between those times.

In the discrete setting, we take $R=\sigma^d$, a hypercell of maximum dimension in our lattice ($d$ is the spacetime dimension). The boundary phase space no longer consists of fields, but of pairs ${\left(\phi_i,\pi_i(\sigma^d)\right)}$ for each vertex $\sigma_i\subset\sigma^d$. The discrete action over $\sigma^d$, $L(\sigma^d)$, is a generating function for a manifold in the boundary phase space: $\left\{\left(\phi_i,\pi_i(\sigma^d)=\frac{\partial L(\sigma^d)}{\partial \phi_i}\right)\right\}$, in agreement with the continuum multisymplectic structure discussed above. Note that in this case, the momentum at a vertex is not unique, but rather depends on the hypercell $\sigma^d$ used to compute it (the same holds true in the continuum: the momentum depends on both the location and the boundary used to define it). The equations of motion (e.g. $\sum_{\sigma^d\supset \sigma_i} \frac{\partial L(\sigma^d)}{\partial \phi_i} =0$) do not guarantee a unique momentum at each vertex, but rather that the sum of momenta defined for each region/boundary containing that vertex vanishes. This guarantees the integrator will be symplectic when stepping in time.

To see this, let vertex $\sigma_i$ be associated with time $t_0$ and define $\sigma^d_+=\{\sigma^d\supset\sigma_i\text{ with } t>t_0\}$ and $\sigma^d_-=\{\sigma^d\supset\sigma_i\text{ with } t<t_0\}$. Then if $\pi_i^+=-\sum_{\sigma^d\in\sigma^d_+}\pi_i(\sigma^d)$ (the minus sign takes care of the orientation for convenience) and $\pi_i^-=\sum_{\sigma^d\in\sigma^d_-}\pi_i(\sigma^d)$, we get the usual gluing of symplectic transformations under time-stepping: $\pi_i^-=\pi_i^+$. Furthermore, this will hold true no matter how we choose to define our time and associated time-stepping (i.e. if there are multiple ways to perform time-stepping in our cellular complex, each of them will be guaranteed to result in symplectic evolution). This argument neglects subtleties that may arise at boundaries or when the number of vertices changes between time slices. To resolve these, a more global perspective is necessary, following for example the presentation in \cite{dittrich2012canonical,dittrich2013constraint,hohn2015canonical}.

In the case of gravity in four dimensions, we are using 1-form fields rather than scalar fields, so the multimomentum is more naturally a 2-form. Additionally, our two primary fields ($e$ and $\omega$) are conjugate to each other (in the sense that the multimomentum of $\omega$ is a function of $e$, while the multimomentum of $e$ vanishes, leaving behind an ${(e,\omega)}$ phase space). All of this is captured by the discrete Eqs.~(\ref{discreteEinsteinEquation}-\ref{cubicLorentzVariation}). The boundary phase space of a cell $\sigma^4$ consists of pairs $(\ell_{ij}^A,\Lambda^A_{ijB})$ for each edge $\sigma_{ij}\subset\sigma^4$ (rather than for each vertex, as in the case of a scalar field). The discrete action $L(\sigma^4)$ is a generating function, which defines momenta conjugate to $\ell_{ij}^A$ and $\Lambda^A_{ijB}$ as the bracketed summands of Eqs.~\eqref{discreteEinsteinEquation} and Eqs.~(\ref{simplicialLorentzVariation}-\ref{cubicLorentzVariation}), respectively. The momentum conjugate to $\Lambda$ is a function of $\ell$, while the initialization of the algorithm (see the end of Sec.~\ref{EOMSection}) ensures the vanishing of the momentum conjugate to $\ell$ on the inner wall of the double wall boundary (i.e. torsion is made to vanish by construction in the $\epsilon$-width cells making up the double wall). The symplectic structure of the time-stepping then ensures the momentum conjugate to $\ell$ vanishes for the entire complex. In this way the discrete multisymplectic structure of our gravitational integrator reproduces the continuous multisymplectic structure of GR.

\section{Conclusion\label{ConclusionSection}}
We have presented a new numerical scheme for general relativity, detailed in Eqs.~(\ref{discreteEinsteinEquation}) and (\ref{simplicialLorentzVariation})-(\ref{cubicLorentzVariation}). This scheme preserves both the (multi)symplectic structure and local Lorentz invariance of the tetrad formulation of GR. Furthermore, its discrete variables have a clear relationship with their continuum counterparts. As such, this scheme holds promise as an integrator for numerical relativity (its structure preservation maintains exact conservation laws and bounded errors in simulation) and for studying the classical limits of certain quantum gravity theories (such as loop quantum gravity, spin foams, etc.). In these roles, the scheme's symplectic structure promises an improvement over non-symplectic finite-difference and spectral methods, while its natural association with continuum variables makes it a more viable alternative to other symplectic approaches to discrete gravity (most notably Regge calculus). In future work, implementations of this algorithm will be needed to demonstrate its practical utility. Furthermore, like Regge calculus, further study is required to incorporate matter into our approach (though the way forward seems clearer).

%In Eqs.~(\ref{discreteEinsteinEquation}-\ref{cubicLorentzVariation}), we have defined a structure-preserving algorithm for the simulation of vacuum numerical relativity that preserves local Lorentz invariance. In future work, a complete implementation of this algorithm will be needed to demonstrate its practical utility. Furthermore, like Regge calculus, further study is required to incorporate matter into our approach.

It may also be of interest to explore the potential union between the algorithm defined here and other structure-preserving discretizations suitable for numerical relativity. For example, it may be useful to explore the relationship between our holonomy-centric approach and the recently developed technique of group-equivariant interpolation in symmetric spaces \cite{gawlik_interpolation_2018,leok_variational_2019}. It may also be useful to compare our effort with finite element cochain complexes suitable for applications in numerical relativity \cite{arnold_complexes_2021}. In this way, the algorithm we have introduced can be an advantageous starting point for explorations into structure-preserving discrete gravity theories.

\section{Acknowledgments}
Thank you to Hong Qin for helpful discussions and encouragement. Thank you to Robert Littlejohn for discussions and inspiration in the lead-up to this work. This work was performed under the auspices
of the U.S. Department of Energy by Lawrence Livermore National Laboratory under contract DE-AC52-07NA27344. This research was further supported by the U.S. Department of Energy Fusion Energy Sciences Postdoctoral Research Program administered by the Oak Ridge Institute for Science and Education (ORISE) for the DOE. ORISE is managed by Oak Ridge Associated Universities (ORAU) under DOE contract number DE-SC0014664. All opinions expressed in this paper are the authors' and do not necessarily reflect the policies and views of DOE, ORAU, or ORISE. A.S.G. further acknowledges the generous support of the Princeton University Charlotte Elizabeth Procter Fellowship.

% Create the reference section using BibTeX: %
\bibliography{allrefs.bib}

\end{document}